\documentclass{ws-procs9x6}

\begin{document}

%
%
\def\D{{\Delta}}

\title{Meson Systems with Ginsparg-Wilson Valence Quarks}

\author{Andr\'{e} Walker-Loud}

\address{Physics Department, University of Maryland,\\
College Park, MD 20742-4111, USA\\
E-mail: walkloud@umd.edu}

\bodymatter

\bigskip
Mixed action (MA) lattice simulations~\cite{Bar:2002nr} are a new variant of lattice regularization which employ different discretization schemes for the \textit{valence} and \textit{sea} fermions.  This allows one to simulate in the chiral regime with numerically expensive chiral fermions~\cite{Kaplan:1992bt} in the valence sector, which satisfy the Ginsparg-Wilson (GW) relation~\cite{Ginsparg:1981bj} while using numerically cheaper fermions which violate chiral symmetry in the sea sector, for example Wilson or staggered fermions.  The chiral extrapolation formulae determined from effective theories extended to include partial quenching (PQ)~\cite{Bernard:1993sv} and lattice spacing $(a)$ artifacts~\cite{Sharpe:1998xm} will generally involve not only the \textit{physical} counterterms of interest, known as Gasser-Leutwyler coefficients in the case of mesons~\cite{Gasser:1984gg}, but will also involve \textit{unphysical} counterterms corresponding to the finite lattice spacing and partial quenching effects.  For MA theories with GW valence fermions, this is not the case; the extrapolation formulae for meson scattering processes through next-to-leading order (NLO) only involve the \textit{physical} counterterms of interest provided one uses a lattice-physical (on-shell) renormalization scheme~\cite{Chen:2005ab,O'Connell:2006sh,Chen:2006wf}.

It has recently been shown~\cite{Chen:2005ab} that the MA $I=2\ \pi\pi$ scattering length, expressed in terms of leading order (LO) parameters has the following form at NLO,
\begin{multline}
       m_{\pi} a_{\pi\pi}^{I=2} = -\frac{m_{uu}^2}{8 \pi f^2} \Bigg\{ 1 
                +\frac{m_{uu}^2}{(4\pi f)^2} \Bigg[
                        4 \ln \left( \frac{m_{uu}^2}{\mu^2} \right) 
                +4 \frac{\tilde{m}_{ju}^2}{m_{uu}^2} \ln \left( \frac{\tilde{m}_{ju}^2}{\mu^2} \right) 
                -1 +\ell^\prime_{\pi\pi}(\mu) 
                \\
%
             	   	-\frac{\tilde{\D}_{ju}^4}{6 m_{uu}^4}
			-\frac{\tilde{\D}_{ju}^2}{m_{uu}^2}\, \ln \left( \frac{m_{uu}^2}{\mu^2} \right)
		\Bigg] 
                + \frac{\tilde{\D}_{ju}^2}{(4\pi f)^2}\, \ell^\prime_{PQ}(\mu) 
                + \frac{a^2}{(4\pi f)^2} \ell^\prime_{a^2}(\mu)
                \Bigg\}\, ,
\label{eq:2seaBare}
\end{multline}
where the \textit{valence-valence} meson mass is $m_{uu}^2 = 2B_0 m_u$, the mixed \textit{valence-sea} meson mass is $\tilde{m}_{ju}^2 = B_0(m_u +m_j) +a^2 \D_\mathrm{Mix}$ and the PQ parameter $\tilde{\D}_{ju}^2 = 2 B_0(m_j -m_u) +a^2 \D_\mathrm{sea}$ (which vanishes in the QCD limit).  This expression depends on three unknown counterterms, only one of which is physical, $\ell_{\pi\pi}^\prime(\mu)$, as well as the mixed meson mass shift, $a^2 \D_\mathrm{Mix}$.  If we express $m_\pi a_{\pi\pi}^{I=2}$ in terms of the lattice-physical parameters, the expression is identical in form to the physical expression up to a known additive shift,
\begin{equation}\label{eq:2seaScattLength}
	m_\pi a_{\pi\pi}^{I=2}= -\frac{m_\pi^2}{8 \pi f_\pi^2} \Bigg\{ 1 
                + \frac{m_\pi^2}{(4\pi f_\pi)^2} \bigg[ 
                        3\ln \left( \frac{m_\pi^2}{\mu^2} \right) 
                        -1 -l_{\pi\pi}^{I=2}(\mu) 
		-\frac{\tilde{\D}_{ju}^4}{6\, m_\pi^4} \bigg]
	\Bigg\}\, ,
\end{equation}
and thus proportional to the \textit{physical} counterterm, $l_{\pi\pi}^{I=2}(\mu)$.  The \textit{unphysical} contribution, $\tilde{\D}_{ju}^4 / 6 m_\pi^4$, turns out to about 10\% of the 
physical NLO contribution~\cite{Chen:2005ab,Chen:2006wf} in the recent lattice simulation of this quantity~\cite{Beane:2005rj}.  

This simplification is understood with the realization that every operator in a NLO MA Lagrangian relevant for external \textit{valence-valence} mesons, incorporating either lattice spacing or PQ effects, amounts to a renormalization of one of the two operators in the LO Lagrangian~\cite{Chen:2006wf}. %
The on-shell renormalization scheme absorbs all \textit{unphysical} counterterm effects into the lattice-physical meson masses and decay constants, removing any explicit dependence upon these operators.  These arguments hold for all meson scattering processes at NLO; non-zero momentum and $N>2$ external mesons.  Finally, the above scattering length, along with $m_K a_{KK}^{I=1}$, $\mu_{K\pi} a_{K\pi}^{I=3/2}$ and $f_K / f_\pi$, two of which have been computed with MA lattice QCD~\cite{Beane:2006kx,Beane:2006gj} share only two linearly independent counterterms, both of which are \textit{physical}, allowing us to make a prediction~\cite{Chen:2006wf} of $m_K a_{KK}^{I=1}$, and thereby test this MA formalism as well as the convergence of $SU(3)$ chiral perturbation theory.

\bibliographystyle{ws-procs9x6}
\bibliography{CD2006}

\end{document}